\documentclass[10pt,letterpaper,twocolumn]{article} 

\usepackage{ol2}
\usepackage[draft,implicit=false]{hyperref}
\usepackage{amsmath}

\usepackage{amsfonts}
\usepackage{subfig}
\usepackage{amssymb}
\usepackage{graphicx}
\usepackage[linesnumbered,ruled,lined]{algorithm2e}
\usepackage{array}
\usepackage{bm}
\usepackage{color}
\usepackage{multirow}
\usepackage{hyperref}
\usepackage{upgreek}

\begin{document}

\twocolumn[ 

\title{Content adaptive sparse illumination for Fourier ptychography}
\vspace{-3mm}
\author{Liheng Bian,$^{1}$ Jinli Suo,$^{1}$ Guohai Situ,$^2$ Guoan Zheng,$^3$ Feng Chen,$^{1}$ and Qionghai Dai$^{1,*}$}

\address{
$^1$Department of Automation, Tsinghua University, China\\
$^2$Shanghai Institute of Optics and Fine Mechanics, Chinese Academy of Sciences, China \\
$^3$Biomedical Engineering $\&$ Electrical and Computer Engineering, University of Connecticut, USA \\
$^*$Corresponding author: qhdai@tsinghua.edu.cn
}

\begin{abstract}Fourier Ptychography (FP) is a recently proposed technique for large field of view and high resolution imaging. Specifically, FP captures a set of low resolution images under angularly varying illuminations and stitches them together in Fourier domain. One of FP's main disadvantages is its long capturing process due to the requisite large number of incident illumination angles. In this letter, utilizing the sparsity of natural images in Fourier domain, we propose a highly efficient method termed as AFP, which applies content adaptive sparse illumination for Fourier ptychography by capturing the most informative parts of the scene's spatial spectrum. We validate the effectiveness and efficiency of the reported framework with both simulations and real experiments. Results show that the proposed AFP could shorten the acquisition time of conventional FP by around 30$\%$-60$\%$.\end{abstract}

\ocis{(110.1758) Computational imaging; (110.3010) Image reconstruction techniques; (170.0180) Microscopy;.}

 ] 

\noindent Fourier Ptychography (FP) is a newly proposed technique for large field of view (FOV) and high resolution (HR) imaging [1-3]. This method sequentially captures a set of low resolution (LR) images describing different spatial spectrum bands of the light field, and these spectrum bands retrieved from captured LR images\cite{Phase_Retrieval} are stitched together in Fourier domain to reconstruct the entire HR spatial spectrum and corresponding HR image.
Recently, Zheng et al. \cite{Nature} extend FP technique to microscopic imaging and propose Fourier Ptychography Microscopy (FPM). FPM assumes that the incident light of a microscope is plane wave, and thus different incident angles result in different shifted versions of the spatial spectrum of the light field through specimen.
Since the whole light field is filtered by the microscope's objective and sampled by CCD, microscopy under changing incident illumination angles results in a set of images corresponding to different spectrum bands of the entire light field. According to \cite{Nature}, it takes the FPM setup about 3 minutes to capture enough images under various illumination angles for reconstructing a gigapixel grayscale image, whose optical resolution is $\sim$0.78 $\mu$m and FOV as $\sim$120 mm$^2$. If one intends to obtain a color image, the capturing time would be multipled. What's more, the calculation time will also increase linearly with the number of captured images.

\begin{figure}[t]
\centering
\centerline{\includegraphics[width=0.9\linewidth]{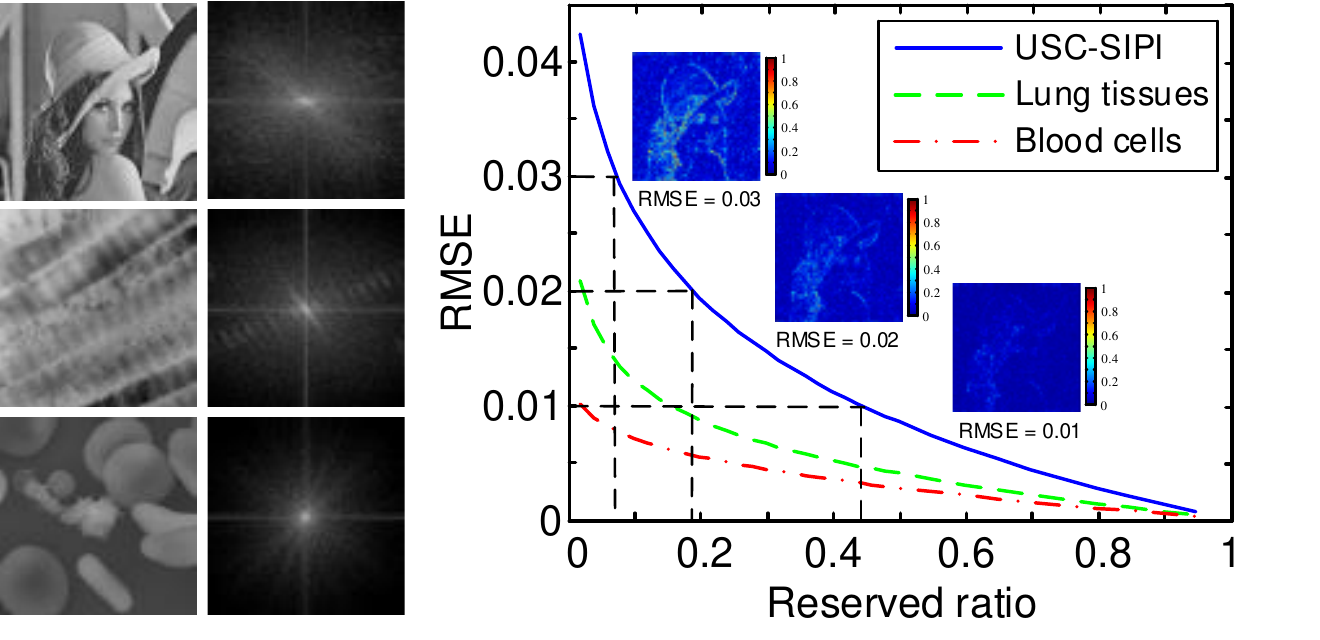}}
\caption{{\bf Illustration of the sparsity prior of natural images in Fourier domain.} The relationship between reserved ratio in Fourier domain and recovered RMSE in spatial domain of three kinds of public images including common natural scenes (USC-SIPI dataset), lung tissues and blood cells (example images and corresponding spatial spectra are shown on the left) is plotted in the right figure. Besides, reconstructed errors of the "Lena" image with different reserved ratios are presented in the figure as well (amplified by 5 times for better visualization).}
\label{fig:Prior}
\vspace{-5mm}
\end{figure}

There have been several studies improving FPM and FP from mainly two aspects. The first aspect is to correct various system aberrations. Bian et al. \cite{Adaptive} derive an adaptive wavefront correction framework for FPM to minimize the error between the captured intensity image set and reconstructed image. This method can correct illumination fluctuation and pupil
aberration effectively, but the global optimization process is time consuming.
For efficiency improvement, a fast embedded pupil function recovery algorithm for FPM is proposed in \cite{Pupil_Function} based on the extended ptychographic iterative engine\cite{ePIE}, which is designed for fast probe function correction in conventional ptychography. The second kind of improvements is to reduce the hardware costs and system's running time. Multiplexing illuminations are applied to FP \cite{Multiplex_1, Multiplex_2} and high quality HR images could be reconstructed using less shots.
Dong et al. \cite{Sparse} investigate the requisite spatial data redundancy of FP and report an FP scheme 
that needs only one shot for each illumination angle instead of multiple differently exposed images. However, this study does not reduce the requisite large number of incident illumination angles, which also leads to long acquisition time of FPM and other FP setups. 

To address the efficiency problem in current FPM, this letter investigates the spectrum redundancy of natural images in Fourier domain and proposes to stitch sparsely sampled but most informative subregions of spatial spectrum to reconstruct the latent HR image. Sparse sampling strategy will decrease the requisite illumination angles during acquisition, and correspondingly decrease the capturing time of FP technique.

To study the redundancy\cite{Prior} in the spatial spectrum of a natural image statistically,  we transform several public image data sets including both macroscopic (USC-SIPI common miscellaneous dataset \cite{Dataset_1}) and microscopic scenes (lung tissues and blood cells \cite{Dataset_2}) to Fourier domain, and then transform them back after truncating small values and calculate root-mean-square error (RMSE) of the reconstructed images.
The relationship between reserved ratio of Fourier coefficients and RMSE is plotted in Fig. {\ref{fig:Prior}}, where we also display the spatial spectrum of several exemplar images for further analysis.
Besides the well known and widely existing sparsity of natural images' Fourier coefficients, 
they also exhibit some other properties. The spatial spectrum is apparently not uniform but exhibits inhomogeneous and monotonously decreasing trends along the radial direction. Besides, the spectrum is of some directionality whose distribution is closely related to the structure of the image. These inspiring observations motivate us to propose the adaptive Fourier ptychography (AFP) which does not exhaustively sample scene's Fourier domain as conventional FP \cite{Phase_FPM}, but utilizes sparse illuminations to adaptively sample the most important bands of the HR spectrum from central low frequency region outward towards the high frequency regions, i.e., in a circle-wise manner.


\begin{figure}[t]
\centering
\centerline{\includegraphics[width=9cm]{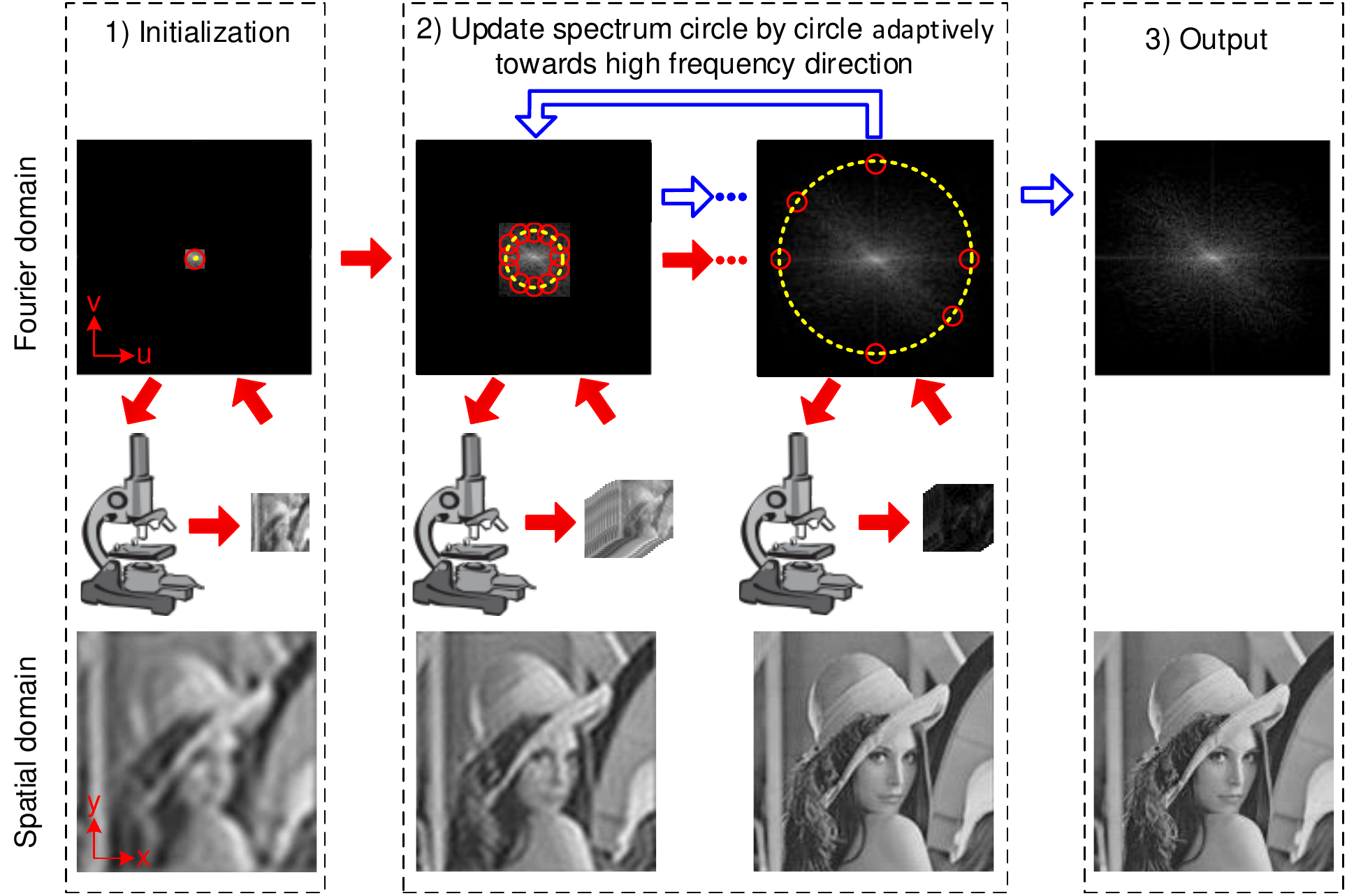}}
\caption{{\bf Flow chart of the proposed APF framework.} After initialization as the left part shows, final results can be exported by iteratively and adaptively updating the HR spectrum (as the blue hollow arrows show) using a small number of images captured in the first iteration (as the red solid arrows show) under different incident illuminations which correspond to the red-labeled spectrum bands shown in the central part.}
\label{fig:FlowChart}
\vspace{-5mm}
\end{figure}

The entire framework of AFP is diagramed in Fig. \ref{fig:FlowChart}. Before detailed explanation of the proposed AFP framework, we 
firstly define a term "potential position", which defines the center of a subregion of a HR spatial spectrum that is potentially to be sampled. 
In addition, considering different pixel numbers between the captured LR images and the recovered image, all the spectra in the entire framework are normalized by dividing corresponding pixel numbers.
What's more, we respectively use x-y and u-v coordinate to describe an image in spatial domain and in Fourier domain. With the above notations, the framework of AFP is detailed as follows.

Firstly, the initial input of the reconstruction framework is set to be the zero-padding extension of spatial spectrum of the LR image shot under normal incident light. Besides, the initial potential position is set to be center of the HR spectrum.

Then we begin loops to update the HR spectrum.
At the start of each loop, we determine potential positions of current loop in the HR Fourier domain by extending the positions of the previous loop towards the high frequency direction with a fixed step size (determined by the user-defined spectrum overlapping ratio), as the yellow dashed line shows in Fig. \ref{fig:FlowChart}.
After getting potential positions in current loop, we decide whether to update the subregion corresponding to each potential position. For each subregion, it will be updated only when it satisfies following two conditions simultaneously: 1) the maximum amplitude of the entries in this subregion is bigger than a manually set threshold, and 2) the closest distance from this subregion's center to other already-updated centers is bigger than the step size. If one subregion is determined to be updated, the LR image under corresponding angular incident light should be captured as the red solid arrows show, and the subregion is then updated in the same way as FP. The correspondence between the subregion's center and the incident angle can be drawn from \cite{Fourier_Optics} as follows. If CCD's sampling rate, namely CCD's pixel size $d$, satisfies the sampling requirement $d\leqslant \frac{\lambda}{2NA}$ \cite{Nature} where $\lambda$ is the wavelength of incident light and $NA$ is the numerical aperture of the microscope's objective, then the correspondence is
\begin{equation}
\Delta u = \frac{dM}{\lambda}\cos(\hat{\alpha})~~\text{and}~~\Delta v = \frac{dN}{\lambda}\cos(\hat{\beta}),
\end{equation}
where $\Delta u$ and $\Delta v$ are the shifts of the subregion's center regarding to the center of the HR spectrum in Fourier domain; $M$ and $N$ are the pixel numbers of the captured LR images along two dimensions; $\hat{\alpha}$ and $\hat{\beta}$ are the incident direction angles regarding to the specimen plane. In addition, the filtering function of the objective is set to be the same as \cite{Nature}, i.e. a circular pupil whose radius is $\frac{NA\times 2\pi}{\lambda}$ in Fourier domain.

After traversing all the potential positions in current loop, the construction process moves to next loop until the potential positions reach the highest frequency of the HR spectrum, i.e., the spectrum boundary.

To improve the recovery accuracy further, users can repeat the above updating iteration in Fourier domain multiple times (experimentally $\sim$3 times are enough for a convergent result, which is the same as FP), as the blue hollow arrows in Fig. \ref{fig:FlowChart} illustrate. There are two worth noting difference between the repetitive iterations and the first one: (i) The previously captured LR images are utilized to avoid repetitive acquisition, i.e., we only adaptively capture LR images under angularly varying illuminations in the first iteration (as the red solid arrows show). (ii) The reconstruction result of the previous iteration instead of the initial image serves as the input of current iteration.

\begin{figure}[t]
\centering
\subfloat[]{
    \includegraphics[width=0.29\textwidth]{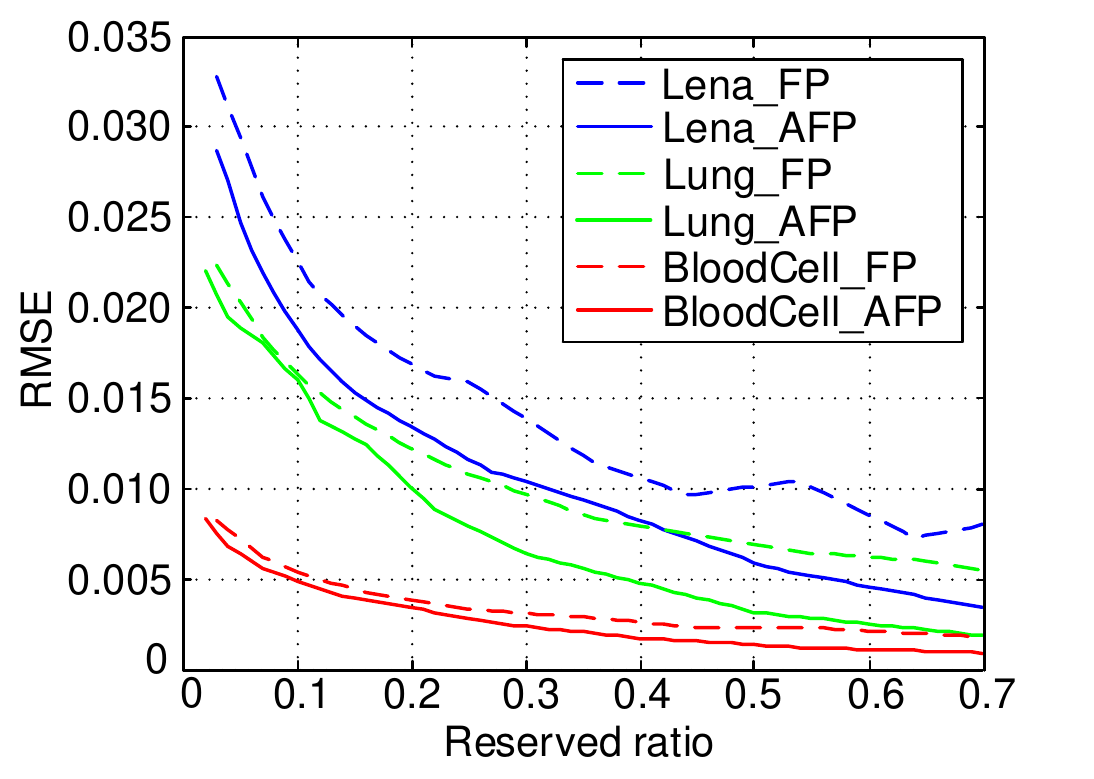}}
\subfloat[]{
    \includegraphics[width=0.17\textwidth]{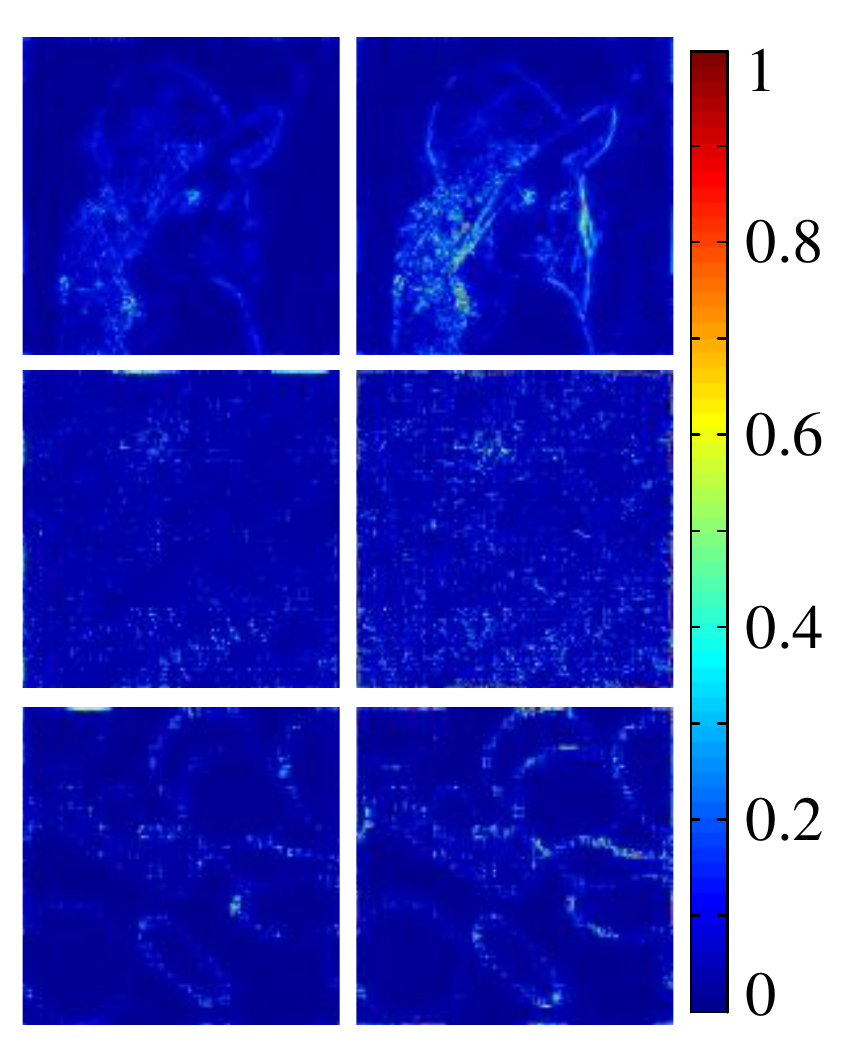}}
\caption{{\bf Comparison of simulation results by AFP and FP on images of different spectrum sparsities and directivities.} (a) plots reconstructed errors of AFP (solid lines) and FP (dashed lines) at different spectrum reserved ratios. The blue, green and red lines respectively correspond to the experiment results conduced on Lena, lung tissue and blood cell image shown in Fig. \ref{fig:Prior}. (b) shows example normalized reconstruction error images of AFP (left column) and FP (right column) with the same reserved ratio (25$\%$) on the three images.}
\label{fig:Simulation}
\vspace{-5mm}
\end{figure}

To measure the advantage of AFP over conventional FP quantitatively, 
here we respectively apply FP and the proposed APF on simulated inputs, and show pros and cons of the two reconstruction methods.
The captured image volume is synthesized in following three steps: we do FFT to the original HR image, and select subregions corresponding to calculated incident angles by multiplying the HR spectrum with an ideal pupil function (all ones in the pupil circle and zeros outside the circle). Then by shifting these subregions to spectrum center and doing inverse FFT, we get complex images in spatial domain. Finally, we remove the phase of these images and retain only their amplitude as LR input images.

In the simulation, LR images' pixel numbers are set to be one tenth of the original HR image along each dimension, and overlapping ratio of spectrum subregions is $65\%$, which is the same as in \cite{Nature}. Besides, the initial HR image is set to be all-one matrix, and the iteration number is set to be 5 times. The user-defined normalized updating threshold ranges from $\frac{10}{256^2}$ to $\frac{100}{256^2}$ with a step as $\frac{10}{256^2}$, which leads to the reserved ration ranging from around 70$\%$ down to around 5$\%$. The simulation results are shown in Fig. \ref{fig:Simulation}. In (a), we compare AFP's reconstruction errors with that of FP on different images and different spectrum reservation ratios. The plots on three kinds of images consistently show that AFP is of less reconstruction errors than FP under the same reserved ratio. In other words, AFP needs less shots than FP to obtain the same recovery precision. Specifically, the amount of saved time is related to the sparsity and directionality of the scene's Fourier coefficients. For Lena and the lung tissue images which contain much high frequency and of strong directionality in Fourier domain, APF could save more than 50$\%$ shots, while for the blood cell image which owns less high frequency and directionality, AFP could save more than 30$\%$ shots. In Fig. \ref{fig:Simulation}(b), three groups of reconstructed error images are represented. The spectrum reserved ratio is set to be the same as 25$\%$, and the errors are normalized by the maximum error value of FP for visualization. From the results we can clearly see that the recovered errors of AFP are much less than that of FP, which means that AFP preserves more crucial image details while of lower noise than FP under the same spectrum reserved ratio.

\begin{figure*}[t]
\centering
\subfloat[]{
    \includegraphics[width=0.3\textwidth,height=0.27\textwidth]{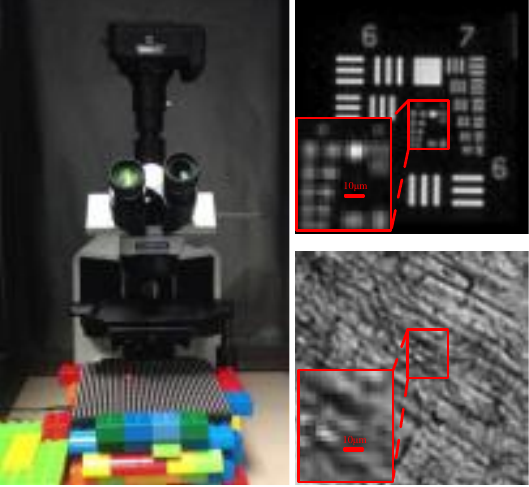}}
\subfloat[]{
    \includegraphics[width=0.3\textwidth,height=0.272\textwidth]{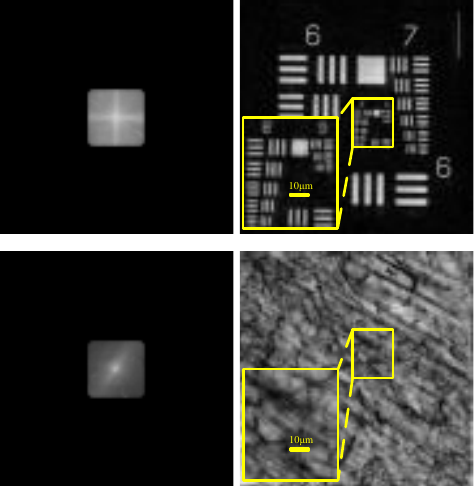}}
\subfloat[]{
    \includegraphics[width=0.3\textwidth,height=0.272\textwidth]{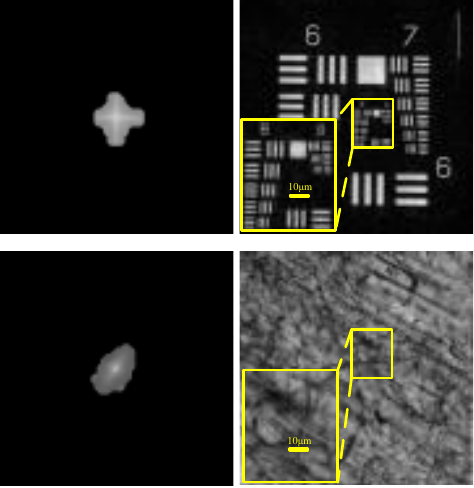}}
\caption{{\bf The prototype setup and real reconstruction results.} (a) shows the prototype microscopy system, as well as two examples of LR images captured under normal illumination including USAF chart and mouse brain slice. (b) presents reconstructed spatial spectra (amplitude) and corresponding HR images by FP utilizing sequential 225 LR images, while the results by the proposed AFP are shown in (c), where the user-defined normalized threshold is set to be $1.5\times 10^{-3}$ (85 LR input images used for the USAF chart and 72 images used for the brain slice).}
\label{fig:Real_Results}
\vspace{-4mm}
\end{figure*}


To further validate the proposed AFP, we build a microscopy setup to capture raw images, as shown in Fig. \ref{fig:Real_Results}(a). The setup parameters are almost the same as the FPM setup in \cite{Nature} except that Olympus BX43 microscope and 2$\times$ objective (NA = 0.08) are utilized for our commercial availability. The LED plate is placed around 8cm under the specimen, and the lateral distance between two adjacent LEDs is 4 mm which is determined by the use-defined spectrum overlapping ratio (35$\%$) and the objective. The central wavelength of incident light is 632nm. The pixel size in the final shot raw images is calculated by dividing the detector's pixel size ($\sim$5.8$\mu$m) with the objective's magnification factor (2) as $\sim$2.9$\mu$m, which is smaller than the maximum pixel size of raw images required by the sampling requirement as $\frac{\lambda}{2NA_{obj}} = 3.95\mu m$.
We capture three LR images for each illumination angle whose exposure time are respectively 0.005s, 0.05s and 0.5s, and then combine them to obtain a high-dynamic range image for subsequent reconstruction.
In addition, the iteration is still set to be 5 times and the embedded pupil function correction technique \cite{Pupil_Function} is utilized in the reconstruction process for smaller reconstruction error.

Results recovered by FP and the proposed AFP are respectively shown in Fig. \ref{fig:Real_Results}(b) and Fig. \ref{fig:Real_Results}(c), where the resolution of LR shot images are $100\times100$, and reconstruction magnification factor is 15. From the figure we can see that AFP works successfully on real captured images including both of the USAF chart and the mouse brain slice (fiber tissue). Specifically, the USAF recovery results show that with 85 LR images AFP could resolve the feature of group 9, element 3 on the USAF target, which achieves the same resolution obtained by FP with 225 LR images. Namely, around 60$\%$ shoots are saved by the proposed AFP method.

\begin{figure}[t]
\centering
\subfloat[]{
    \includegraphics[width=0.14\textwidth,height=0.25\textwidth]{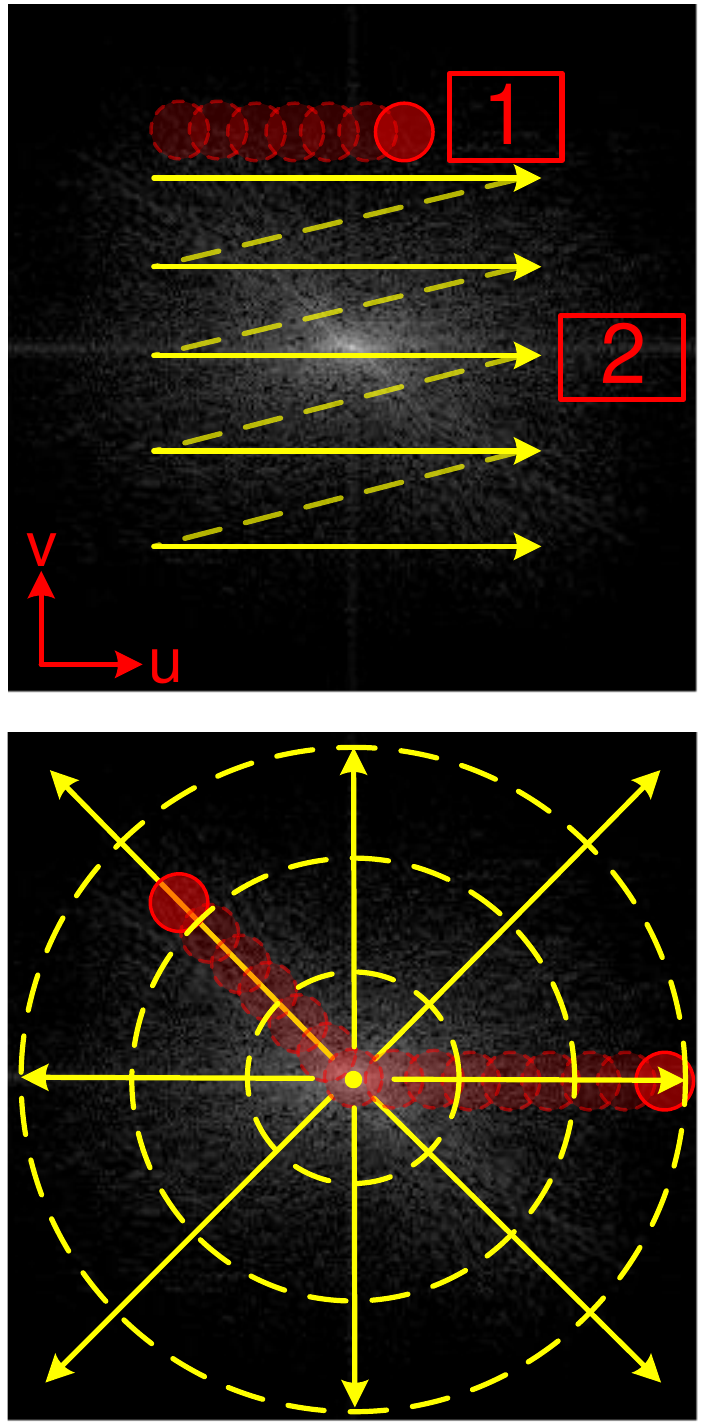}}
\subfloat[]{
    \includegraphics[width=0.14\textwidth,height=0.25\textwidth]{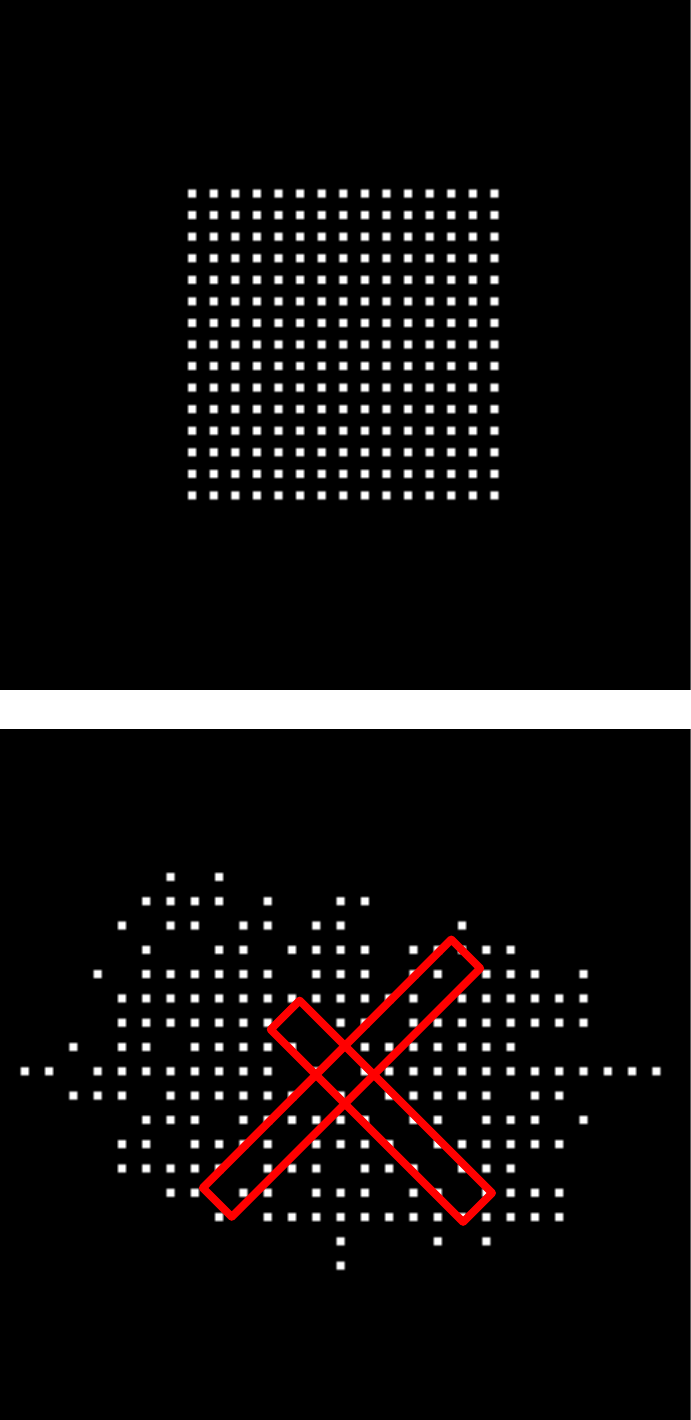}}
\subfloat[]{
    \includegraphics[width=0.14\textwidth,height=0.25\textwidth]{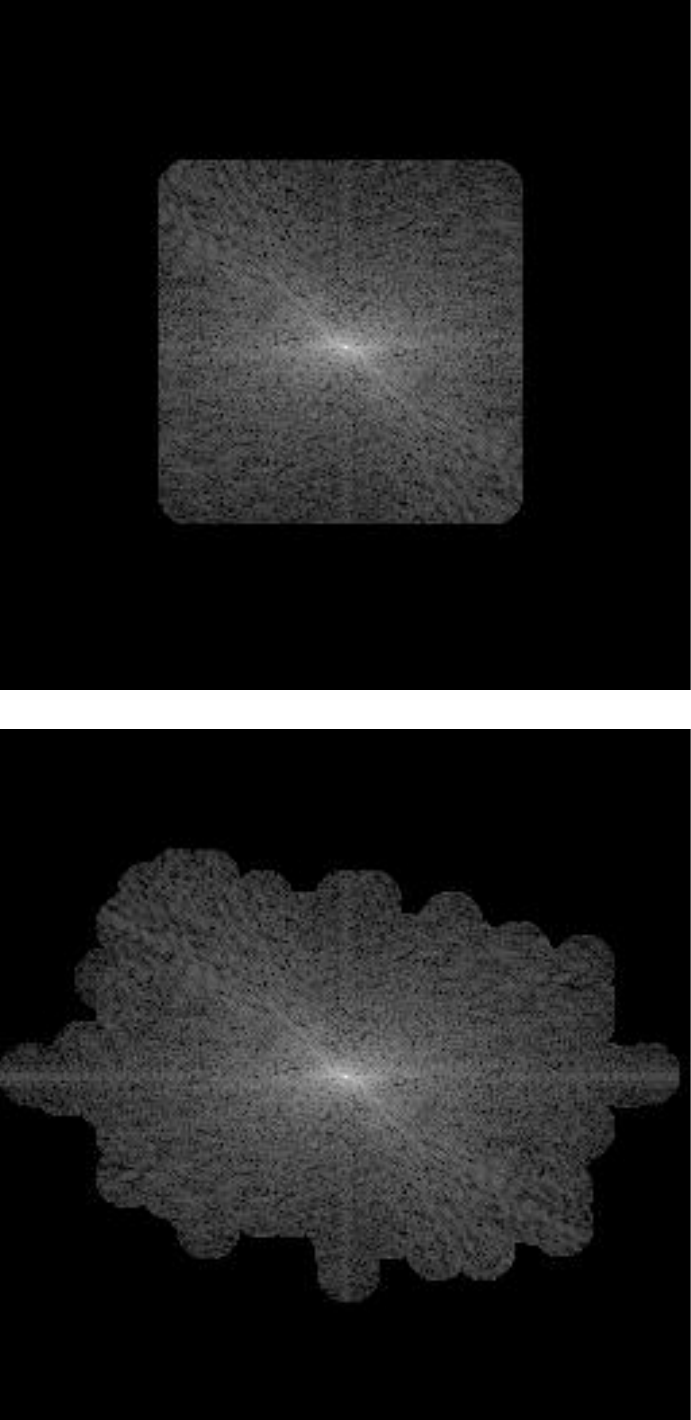}}
\caption{{\bf Comparison of Fourier domain sampling sequence beween FP (top row) and AFP (bottom row) on the Lena image.} (a) shows the theoretical sampling sequences of FP and AFP. (b) and (c) respectively present the simulated sampling central positions and recovered spectra of Lena image by FP and AFP under the same sampling ratio as 25$\%$.}
\label{fig:Comparison}
\vspace{-5mm}
\end{figure}

Both the above simulation and real-shot experiments indicate that AFP could significantly shorten the acquisition time compared to FP. To explain the source of AFP's efficiency advantages, we illustrate its sampling strategy in parallel with that of conventional FP in Fig. \ref{fig:Comparison}. As shown in (a), FP captures all the images corresponding to the central part of HR spectrum exhaustively and then updates the spectrum sequentially. There are two main disadvantages of FP's sampling method: 1) Some areas of small amplitudes such as the '1' labelled region in (a) are updated, but this improves little reconstruction performance, so capturing their corresponding images is inefficient. 2) Because FP only samples the central part of the whole spectrum with a light NA as $\sim$0.5 \cite{Nature}, some important high frequencies owning large amplitudes are neglected, such as the '2' labeled area.

Instead, our proposed AFP adaptively updates the spectrum subregions in a circle-wise manner from low frequency to high frequency. In each circle it only updates areas of large spectrum amplitudes. For example, as Fig. \ref{fig:Comparison}(b) and (c) show, the horizonal outward regions of bigger amplitudes are updated selectively, while the regions of smaller amplitudes (such as the top and bottom high frequency regions, as well as some sporadic low frequency regions such as the red box labeled regions) are neglected. To sum up, benefiting from the adaptive updating scheme with highly efficient implementation, AFP can capture the most important information in Fourier domain and thus can save the acquisition time significantly, especially when object images own stronger directivity and thus more sparse spatial spectra.

We thank Xuemei Hu, Dongsheng An, Jingyu Lin, Weixin Jiang and Yulun Zhang for valuable discussions. This work is supported by the National Nature Science Foundation of China under grant 61327902, 61377005 and 61120106003, as well as the Recruitment Program of Global Youth Experts.
\vspace{-5mm}

\clearpage

\section*{Informational Fourth Page}

\end{document}